\documentclass[12pt,preprint]{aastex}

\begin{document}
\shorttitle{\sc THE ABUNDANCE SPREAD IN BO\"{O}TES I }
\shortauthors{NORRIS ET Al.}

\newcommand{\boo} {Bo\"{o}tes~I}
\newcommand{\WAD} {W$_{\rm AD}$}
\newcommand{\Wtwo} {W$_{8542}$}
\newcommand{\Wthree} {W$_{8662}$}
\newcommand{\VHB} {V$_{\rm HB}$}
\newcommand{\kms} {km s$^{-1}$} 
\title{THE ABUNDANCE SPREAD IN THE BO\"{O}TES~I DWARF SPHEROIDAL GALAXY}

\author
{John E. Norris\altaffilmark{1}, Gerard Gilmore\altaffilmark{2}, Rosemary F.G. Wyse\altaffilmark{3}, Mark I. Wilkinson\altaffilmark{2,4}, V. Belokurov\altaffilmark{2},  N. Wyn Evans\altaffilmark{2}, Daniel B. Zucker\altaffilmark{2}}

\altaffiltext{1}{Research School of Astronomy \& Astrophysics, The Australian National University, Mount Stromlo Observatory, Cotter Road, Weston, ACT 2611, Australia; email: jen@mso.anu.edu.au}
\altaffiltext{2}{Institute of Astronomy, University of Cambridge, Madingley Road, Cambridge CB3 0HA, UK}
\altaffiltext{3}{The Johns Hopkins University, Department of Physics \& Astronomy, 3900 N.~Charles Street,  Baltimore, MD 21218, USA}
\altaffiltext{4}{Department of Physics and Astronomy, University of Leicester, University Road, Leicester, LE1 7RH, UK.} 

\begin{abstract}

We present medium-resolution spectra of 16 radial velocity red-giant
members of the low-luminosity {\boo} dwarf spheroidal (dSph) galaxy
that have sufficient S/N for abundance determination, based on the
strength of the Ca II K line.  Assuming [Ca/Fe] $\sim$~0.3, the
abundance range in the sample is $\Delta$[Fe/H] $\sim$~1.7 dex, with
one star having [Fe/H = --3.4.  The dispersion is $\sigma$([Fe/H]) =
 0.45 $\pm$ 0.08  -- similar to those of the
Galaxy's more luminous dSph systems and $\omega$ Centauri.  This
suggests that the large mass ($\gtrsim$~10$^{7}~M_{\odot}$) normally
assumed to foster self-enrichment and the production of chemical
abundance spreads was provided by the non-baryonic material in
{\boo}.\\

\end{abstract}

 
\section{INTRODUCTION}

The Galaxy's dSph satellites continue to play a challenging and
pivotal role in our understanding of the $\Lambda$CDM paradigm and the
manner in which the Milky Way formed.  Not only do there appear to be
too few of these systems in comparison with the predictions of CDM
satellite dark halos (Klypin et al. 1999; Moore et al. 1999),  but for
those who see the dSphs as possible building blocks for the Galaxy's halo  (e.g. Carollo et al. 2007), their stellar populations pose
problems.  Not only are their age distributions too young (Unavane et
al. 1996), their chemistry, at first blush, is wrong, at least as far
as e.g. [$\alpha$/Fe] is concerned (Venn et al. 2004,  and references
therein  ).  The report by Helmi et al. (2006) of a dearth of stars with
[Fe/H] $<$ --3.0 in the more luminous dSphs, if correct, offers no
comfort to those who seek the origin of the now well-established
Galactic extremely metal-poor stars having --4.0 $<$[Fe/H] $<$ --3.0.

The recent discovery of very low luminosity dSph galaxies, down to
M$_{\rm V}$ $\sim$~--4 (e.g. Belokurov et al. 2006, 2007), but with
associated non-baryonic masses of 10$^{7}$~M$_{\odot}$ within the
extent of the stellar distribution (e.g. Gilmore et al. 2007) hints
there is still much to be understood before we have a complete
understanding of the process of the formation of our Galaxy.

An important feature of the more luminous of the Milky Way's dSphs is
that there exists a well-established range in metallicity within each
system, with dispersions of order 0.3--0.5 dex (e.g. Helmi et
al. 2006)\footnote {Kirby et al. (2008) have very recently reported
abundance spreads in several of the Galaxy's less luminous dSph
systems.}.  The purpose of this Letter is to present evidence that
similar behavior pertains to one of the less luminous systems, {\boo},
with M$_{\rm V}$ = --5.8 (Belokurov et al. 2006).  We shall argue that
{\boo} exhibits a range in metallicity of $\sim$~1.7 dex,
$\sigma$([Fe/H]) = 0.45, and stars as metal-poor as
[Fe/H] = --3.4.

\section{SPECTROGRAPHIC OBSERVATIONS AND DATA REDUCTION}

Spectra of $\sim$~330 {\boo} candidate members were obtained with the
Anglo-Australian Telescope/ AAOmega  fiber-fed   spectrograph (http://www.aao.gov.au/local/ www/aaomega/) combination during 2006 May
23--29 and 2007 April 18--20.  In the blue, the spectra have
resolution R = 5000, and cover the wavelength range 3850--4540 {\AA},
while in the red the corresponding numbers are 14000 and 8340--8840
{\AA}, respectively.

Candidates were selected from the SDSS DR4 data set, using the color
magnitude diagram selection mask illustrated in Figure 2 of Belokurov
et al (2006).  Stars in the magnitude range 17 $<$ g $<$ 20.5 were
selected for observation with AAOmega, with a deliberate effort to
select candidates up to one degree from the galaxy centre, four
half-light radii, given the galaxy's apparently distorted and
elliptical image.  Only limited observational material was obtained in
2006.  Deeper data, using essentially the same target selection, were
obtained in 2007.  The 2006 run was the first major visitor use of the
new AAOmega facility.  These data sets were used to optimise and
enhance the data reduction system, 2dfdr  (see
http://www.aao.gov.au/AAO/2df/software.html\#2dfdr)  , in collaboration
with local expertise, leading to its current excellent performance.
Final data calibration and reduction used what is now the public 2dfdr
system.

\section{ANALYSIS}

\subsection{Radial Velocities}

Heliocentric radial velocities were determined using the {\tt HCROSS}
routine of the {\tt FIGARO} package  (see
http://www.starlink.rl.ac.uk/star/docs/sun86.htx/ node425.html)  .  This
performs a cross-correlation between the program stellar spectrum and
a template, to obtain the relative radial velocity.  An associated
confidence level and formal error are estimated and we accept only
those with {\tt confidence = 1} (see Heavens 1993). We excised the
strong Ca II H \& K lines, although this made an insignificant
difference in most cases.  Two template spectra were used, of twilight
sky and a G-giant star, both obtained during the same observing run as
the {\boo} candidates.  The velocities from the two templates were in
general consistent within the errors and here we report those using
twilight sky as template.  We thus obtained velocities for 98 objects,
for which our mean formal error is 3.6~{\kms}.  The internal accuracy
on one observation, from repeat observations of 4 stars from 2006 and
2007, is 7~{\kms} with a mean offset of $-1.5$~{\kms}.  Our external
errors may be estimated by comparison with Martin et al.~(2007).  For
5 stars in common, observed by us in 2007, the mean offset is
4.6~{\kms}, with $\sigma$~=~1.8~{\kms}.

\subsection{Selection of {\boo} Members}

For the purposes of the present work we choose to consider objects for
abundance analysis only if their spectra have net counts greater than
$\sim$~200 per 0.34 {\AA} pixel at 4150 {\AA} and lie within the
radial velocity range 90--115 km s$^{-1}$.  The first criterion is set
to obtain a reliable abundance estimate based on the Ca II K 3933
{\AA} line, while the second admits objects within $\pm$~1.8$\sigma$
of the galaxy's systemic velocity (we adopt a velocity dispersion of
7~{\kms}).  Data for the 16 objects within our sample that satisfy
these criteria are presented in Table~1, where columns (1)--(5)
contain identification, coordinates, radial distance from galactic
center, and heliocentric radial velocity, (6)--(7) present SDSS colors
g$_{0}$ and (g--r)$_{0}$ (we adopt E(B--V) = 0.02, following Belokurov
et al. 2006), and column (8) contains the net counts obtained per 0.34
{\AA} pixel at 4150 {\AA}.  For four objects, data are available from
both the 2006 and 2007 sessions -- in these cases the second entry in
the table refers to data obtained in 2006.

We believe our sample has very small field contamination.  In velocity
windows at lower (65--90 km s$^{-1}$) and higher (115--140 km
s$^{-1}$) values, and for our count restriction, we find a total of 4
objects (one clearly too strong-lined to be a member).  Standard
Galactic components of thin disk, thick disk and halo have predicted
mean velocities\footnote{See erratum in the appendix.} in this line-of-sight, for a typical distance of dwarf
stars of 3 kpc, of --5, --25, and --130 km~s$^{-1}$, respectively,
well-separated from {\boo}.

\subsection{Chemical Abundances from the Ca II K line}

For the putative {\boo} members in Table~1 we have measured the Ca II
K line-strength index, K$^{\prime}$, and the G band index,
G$^{\prime}$, defined by Beers et al. (1999), which we present in
columns (9) and (10) of Table~1.  Beers et al. provide a formalism,
based on Galactic field stars and globular clusters that permits one
to estimate [Fe/H] given K$^{\prime}$ and (B--V)$_{0}$,
 well-calibrated down to [Fe/H] = --4.0,  which we adopt here.  The
reader should be aware that the implicit assumption, that the same
[Ca/Fe] vs [Fe/H] relationship applies within both the metal-poor
Galactic objects used for the calibration and the dSph satellites, is
not consistent with high-resolution abundance studies (e.g. Venn et
al. 2004) at, say, the 0.3 dex level for the dSph systems studied to
date.  Here we shall cite [Fe/H] values obtained from the
Beers et al. calibration, assuming [Ca/Fe] $\sim$~0.3;
allowance for a putative spread in [Ca/Fe] in {\boo} would increase
any inferred dispersion of [Fe/H].    To use the Beers et
al. formulation we have obtained B--V values using the transformation
(B--V)$_{0}$ = 1.197$\times$(g--r)$_{0}$+0.049 valid on the range 0.70
$<$ B--V $<$ 1.3, determined from observations of the red giants in
the Galactic globular clusters M13 and NGC 2419\footnote{Based on data
kindly made available to us by H. L. Morrison (B--V and g--r
photometry of Stetson (2008,
http://www1.cadc-ccda.hia-iha.nrc-cnrc.gc.ca/community/STETSON/standards/)
and An et al. (2008), and cluster membership for NGC 2419 from
Peterson (1985), Shetrone et al. (2001), and Suntzeff et al. (1988)).
We prefer this calibration to those of Lupton (2005),
http://www.sdss.org/dr5/algorithms/sdssUBVRITransform.html) and Zhao
\& Newberg (2006), from stars in the general field and metal-poor
objects having (B--V) $<$ 0.70, respectively.}.

The resulting [Fe/H] values are presented in column (11) of
Table~1.  We find $\langle$[Fe/H]$\rangle$ =~--2.51~$\pm$~0.13, which
agrees well with Mu\~{n}oz et al. (2006, [Fe/H] =~$-2.5$), Martin et
al. (2007, [Fe/H] =~--2.1), and Feltzing et
al. (http://www.mpa-garching.mpg.de/~garcon08/program.html, [Fe/H]
=~--2.7).  Small sample statistics (Keeping 1962) of the abundance
differences for the four stars in Table~1 observed in both 2006 and
2007 yields a standard deviation of a single observation of 0.19
dex\footnote{ A second estimate of our internal accuracy
comes from $\sim$~10,000 dwarfs observed during our 2006 session (and
analyzed with similar techniques), which serendipitously included 16
objects with [Fe/H] $<$ --1.5, and g $>$ 18.0, from SSDS DR6 (see Lee
et al. 2008).  The dispersion of abundance differences between the two
sub-samples (over 8 independent pointings) is 0.23 dex.},
while comparison of results for 7 stars in common with Feltzing et
al. (assuming equal errors in their work and ours) yields 0.35 dex.
Further consideration of Table~1 reveals the principal result of our
investigation: among the 16 putative galaxy members there is a spread
of $\Delta$[Fe/H] $\sim$~1.7 -- considerably larger than expected from
the above measurement accuracies.  The abundance dispersion for our
2007 data, corrected for measurement error of 0.19~(0.35) dex is
0.49~(0.40): we adopt $\sigma$[Fe/H]~=~0.45~$\pm$~0.08 for
{\boo}.

The large range in abundance can also be appreciated in Figure 1,
which presents the spectra for 12 stars in Table~1.  Figure 2 compares
the most metal-poor object in our sample (Boo--1137, with [Fe/H] =
--3.4) with classical metal-poor stars in the range --4.2 $<$ [Fe/H]
$<$ --2.5. The low metallicity of Boo--1137 is
well-confirmed.\footnote {Boo-1137 lies 24$\arcmin$ from the center
of the system, well outside the half-light radius of 13$\arcmin$ (but
along the major axis of the galaxy's elongation).  Given the rareness
of objects with [Fe/H] $<$ --3.0, however, it is very likely to be a
member.  From the HK survey for metal-poor stars, Beers et al.(1992)
find 5 stars with [Fe/H] $<$ --3.0, V $<$ 14, and B--V $>$ 0.7, over
2300 deg$^{2}$.  For the HES, Christlieb (2008, priv. comm.)  finds 9
stars with B $<$ 14.5 and B--V $>$ 0.7, over 7000 deg$^{2}$.  We estimate one
should expect only $\sim$~0.02 halo giants having [Fe/H] $<$ --3.0
within the central 30$\arcmin$ of {\boo}.}

\subsection{The Calcium Triplet}

To complement the Ca II K line data we present equivalent widths
of the Ca II triplet 8542 {\AA} line in column (12) of Table 1.  (Poor
sky subtraction prevents measurement of the triplet 8662 {\AA} line.)
There is a clear and strong correlation between the K line and triplet
line data in Table~1.  The question that needs to be addressed is
whether the triplet data support the K line abundances.

To our knowledge, all abundance calibrations based on the triplet
follow Armandroff \& Da Costa (1991, hereafter AD) and are anchored to
the abundances of Galactic globular clusters, none of which has
[Fe/H] $<$ --2.5.  (This method also assumes [Ca/Fe] $\sim$~0.3.)
Extrapolation is thus necessary when using the triplet for the
$\sim$~half of the objects in Table~1 having [Fe/H] $<$ --2.5.  To make
progress we proceed as follows. Based on spectra of standard globular
cluster giants with [Fe/H] $<$ --1.4 described by Norris et
al. (1996), we transform {\Wtwo} to the AD triplet index {\WAD} =
{\Wtwo}+{\Wthree} using the relationship {\WAD} =
2.34$\times${\Wtwo}. The pivotal diagram of AD is their plot of {\WAD}
vs V--{\VHB} (V is the star's magnitude and {\VHB} refers to the
horizontal branch of the cluster under consideration).  To obtain V we
transform the g$_{0}$ and (g--r)$_{0}$ values in Table~1 following
Lupton (2005, op.cit.)
For {\boo}, {\VHB} = 19.55 (Dall'Ora et al. 2006).  All but two of the
objects in Table~1 fall below the M15 ([Fe/H] = --2.2) data of AD
(their Figure 3) in their uncalibrated region, reasonably consistent
with only four stars in our Table~1 having [Fe/H] $>$ --2.2.

AD also define the reduced index W$^{\prime}$ = {\Wtwo}+{\Wthree}
+0.619$\times$(V--{\VHB}), and determine the calibration [Fe/H] =
[Fe/H]$_{\rm AD}$ $\equiv$ 0.326W$^{\prime}$--2.706, valid for 1.6 $<$
W$^{\prime}$ $<$ 4.2 (--2.2 $<$ [Fe/H] $<$ --1.3).  Using these
relationships for the data in Table~1, we obtain
$\langle$[Fe/H]$\rangle$$_{\rm AD}$ = --2.32 and $\sigma$[Fe/H]$_{\rm
AD}$ = 0.20, somewhat different from the values obtained from the Ca
II K line.  For Boo--1137, with [Fe/H]$_{\rm Ca~II~K}$ = --3.4, we
find [Fe/H]$_{\rm Ca~II~triplet}$ = --2.7.

We found no model-atmosphere calibrations of the triplet in the
literature for giants with [Fe/H] $<$ --2.5.  This is perhaps not
surprising: in this regime the triplet is relatively weak,
core-dominated, and challenging to model atmosphere analysis.  Using
the spectrum synthesis code MOOG of Sneden (2008,
http://verdi.as.utexas.edu/moog.html), models of Kurucz (1993), atomic
data from VALD (http://www.astro.uu.se/$\sim$vald/), and a range of
van der Waals damping values for the triplet, together with T$_{\rm
eff}$, log~$g$, and colors from the Yale-Yonsei isochrones
(http://www.astro.yale.edu/demarque/yyiso.html) we computed spectra
for [Fe/H] = --1.5 to --4.0, and plotted isoabundant loci in the
{\WAD} vs (V--{\VHB}) plane.  The agreement between observation and
theory leaves something to be desired: while the observed loci are
essentially parallel, as are the theoretical ones, the latter have a
shallower slope.  Most importantly, the theoretical lines move closer
together at lowest abundance.  We use these results to indicate,
approximately, the improvement we believe necessary to extrapolate the
AD relationship.  For [Fe/H] $>$ --2.5, we adopt [Fe/H] = [Fe/H]$_{\rm
AD}$ = 0.326$\times$W$^{\prime}$--2.706, and, for --4.0 $<$ [Fe/H] $<$
--2.5, [Fe/H] = [Fe/H]$_{\rm AD}$ --1.03 + 0.83$\times$W$^{\prime}$
--0.15$\times$W$^{\prime}$$^{2}$.  Then for the data in Table~1, we
find $\langle$[Fe/H]$\rangle$ = --2.64, $\sigma$[Fe/H] = 0.45, and
[Fe/H]$_{\rm Boo-1137}$ = --3.8.

\section{DISCUSSION}

Despite its small baryonic mass, {\boo} has an abundance dispersion
very comparable with those of the more massive dSphs ($\sigma$[Fe/H] =
0.3--0.6, e.g. Helmi et al. 2006) and the most massive
(3$\times$10$^{6}$~M$_{\odot}$) Galactic globular cluster $\omega$ Cen
($\sigma$[Fe/H] = 0.3, Norris et al. 1996).  With a current stellar
mass of $\sim$~4$\times$10$^{4}$ M$_{\odot}$ (Fellhauer et al. 2008),
however, the chemical inhomegeneity of {\boo} stands in stark contrast
against the heavy-element homogeneity of Galactic globular clusters of
similar stellar mass.  The obvious solution is that {\boo} sits within
the deep potential well of a dark halo mass -- of order 10$^{7-8}$
M$_{\odot}$.  Given that chemical inhomogeneity within globular
clusters (which contain no dark mass) appears to occur only in massive
objects such as $\omega$ Cen, it seems reasonable to assume that a
requirement for the creation of heavy-element chemical inhomogeneity
is that {\boo} should have a potential well deep enough to retain the
ejecta of Type II supernovae for periods sufficiently long for them to
be incorporated into subsequent stellar generations.

There is a test of this: self-enrichment in the lowest luminosity
systems will be greatly affected by sampling noise. As demonstrated by
the dynamical modelling of Fellhauer et al. (2008), if {\boo} ever had
a dark matter halo, it still has it and has lost very little of its
stellar mass.  That is to say, {\boo} has always had relatively little
stellar material.  If one normalizes abundance dispersion with respect
to luminosity or current baryonic mass, the specific abundance
dispersion of {\boo} is some 10--100 times larger than that of the
more luminous dSph. Not many supernovae are needed to chemically
enrich, say, 10$^{5}$ M$_{\odot}$ of pristine material to an abundance
of [Fe/H] = --2.5.  One should expect very specific and different
element patterns in the stars of {\boo} directly from this very small
number of progenitors.  One may be looking at the yields of individual
supernovae in the abundances of these stars.\vspace{-3mm}

\acknowledgements

The authors gratefully acknowledge the contributions of the AAOmega
project team, in particular Rob Sharp, during this investigation.
We thank  H.L. Morrison for generous instruction in (B--V, g--r) --
transformations.

\clearpage
\appendix
\section{Erratum}

An unfortunate confusion over coordinates
resulted in incorrect predictions for the velocity distributions of
foreground Galactic stars in \S3.2.  Instead of the quoted negative
velocities, the correct predicted velocity distribution peaks close to
zero heliocentric velocity for all Galactic populations, with only a
small contribution (around 3\% of the sample) in the velocity range of
the Bootes I system.  None of the analysis or conclusions of the paper
is affected by this error.

\clearpage

\begin{deluxetable}{rccrrccccccc}
\tabletypesize{\scriptsize}
\tablecaption{OBSERVATIONAL DATA AND DERIVED PARAMETERS FOR BO\"{O}TES I\vspace{0mm}}
\tablewidth{0pt}
\tablehead{
\colhead {Star} & {RA}        & {Dec}    & {r}      & {V$_{r}$}      & {g$_{0}$}  & {(g-r)$_{0}$}   & {C4150} & {K$^{\prime}$}      & {G$^{\prime}$}      & {[Fe/H]} & {W8552}   \\
         {}     & {(2000)}    & {(2000)} & {($\arcmin$)} & {(kms$^{-1}$)} & {}         & {}              & {}      & {({\AA})} & {({\AA})} & {}       & {({\AA})} \\ 
         {(1)}  & {(2)}       & {(3)}    & {(4)}    & {(5)}          & {(6)}      & {(7)}           & {(8)}   & {(9)}     & {(10)}    & {(11)}   & {(12)}      
}
\startdata
7    & 13 59 35.53 & +14 20 23.7 & 12.1 &  104 & 18.31 & 0.70 & 609 & 7.49 & 2.01 & $-$2.32 & 1.35 \\   
8    & 13 59 38.62 & +14 19 15.9 & 12.6 &  109 & 19.03 & 0.56 & 345 & 4.53 & 3.25 & $-$2.75 & 0.84 \\   
9    & 13 59 48.81 & +14 19 42.9 & 11.1 &  108 & 17.92 & 0.80 & 772 & 6.62 & 1.99 & $-$2.67 & 1.16 \\   
33   & 14 00 11.73 & +14 25 01.4 &  5.2 &  105 & 18.23 & 0.74 & 321 & 5.17 & 5.28 & $-$2.96 & 1.20 \\   
34   & 14 00 21.11 & +14 17 27.9 & 13.1 &  104 & 18.74 & 0.65 & 313 & 4.94 & 2.66 & $-$3.10 & 1.20 \\   
41   & 14 00 25.83 & +14 26 07.6 &  6.2 &   98 & 18.38 & 0.70 & 239 & 8.13 & 4.29 & $-$2.03 & 2.17 \\   
78   & 14 00 14.73 & +14 13 13.9 & 16.9 &  102 & 19.30 & 0.64 & 237 & 6.10 & 2.24 & $-$2.46 & 1.14 \\   
     &             &             &      &      &       &      & 202 & 4.88 & 1.48 & $-$3.02 & 1.25 \\   
94   & 14 00 31.51 & +14 34 03.6 &  7.4 &   93 & 17.53 & 0.90 & 538 & 6.63 & 0.14 & $-$2.79 & 1.24 \\   
117  & 14 00 10.49 & +14 31 45.5 &  2.1 &   94 & 18.19 & 0.74 & 283 & 8.78 & 3.46 & $-$1.72 & 1.46 \\   
121  & 14 00 36.52 & +14 39 27.3 & 12.0 &  113 & 17.92 & 0.83 & 237 & 7.62 & 3.23 & $-$2.37 & 1.51 \\   
     &             &             &      &      &       &      & 387 & 7.43 & 3.14 & $-$2.46 & 1.54 \\   
127  & 14 00 14.57 & +14 35 52.7 &  6.2 &   99 & 18.15 & 0.76 & 236 & 9.47 & 4.72 & $-$1.49 & 1.67 \\   
     &             &             &      &      &       &      & 346 & 9.36 & 3.96 & $-$1.49 & 1.86 \\   
130  & 13 59 48.98 & +14 30 06.2 &  4.1 &  110 & 18.21 & 0.70 & 337 & 6.60 & 2.13 & $-$2.55 & 1.22 \\   
911  & 14 00 01.08 & +14 36 51.5 &  7.0 &   97 & 17.94 & 0.75 & 237 & 8.47 & 3.44 & $-$1.98 & 1.51 \\   
980  & 13 59 12.68 & +13 42 55.8 & 48.8 &  104 & 18.51 & 0.61 & 396 & 4.14 & 0.82 & $-$3.09 & 0.86 \\   
1069 & 13 58 53.22 & +14 06 57.8 & 29.0 &  111 & 19.05 & 0.56 & 490 & 5.57 & 1.27 & $-$2.51 & 0.96 \\   
     &             &             &      &      &       &      & 200 & 3.80 & 1.52 & $-$2.91 &  ... \\   
1137 & 13 58 33.82 & +14 21 08.5 & 24.0 &  112 & 18.11 & 0.73 & 859 & 3.06 & 1.09 & $-$3.45 & 0.66 \\   
\enddata
\end{deluxetable}

\clearpage
\begin{figure}[htbp]
\vspace{1cm}
\begin{center}
\includegraphics[width=12.5cm]{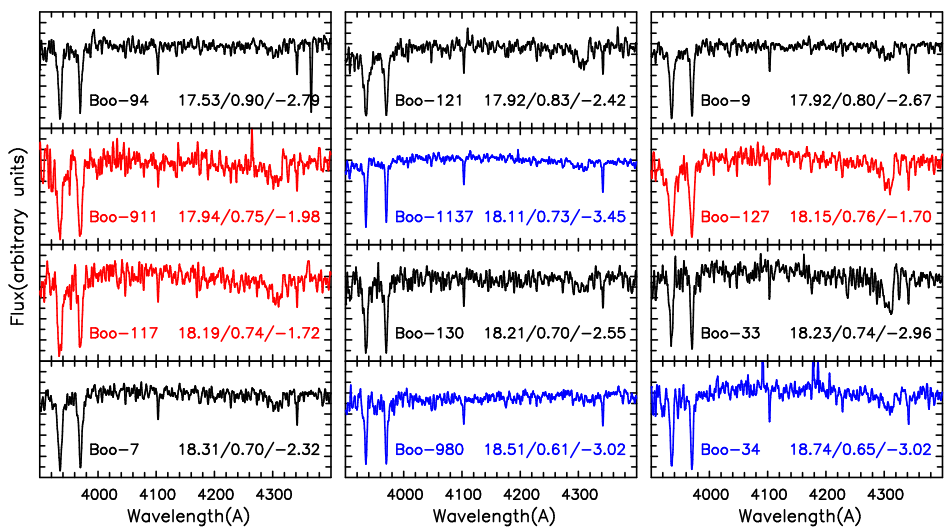}
\figurenum{1}

  \caption{Spectra of 12 {\boo} red giants from Table~1
  (continuum-normalized and broadened to resolution of FWHM =
  2.5~{\AA}).  Panels contain star name and g$_{0}$/(g-r)$_{0}$/[Fe/H].
  Objects with [Fe/H] $<$ --3.0 are presented in blue, while red is
  used for those with [Fe/H] $>$ --2.0.}

\end{center}
\end{figure}

\clearpage
\begin{figure}[htbp]
\begin{center}
\includegraphics[width=6cm]{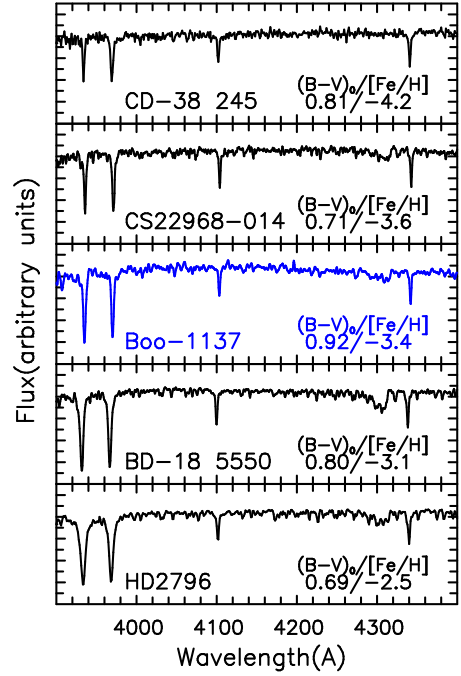}
\end{center}
\vspace{-5mm}
\figurenum{2}

\caption{Spectra of Boo--1137 ([Fe/H] = --3.4) and extremely
  metal-poor field giants of similar temperature (from ANU's 2.3m
  telescope; [Fe/H] and (B-V)$_{0}$ from Cayrel et al. (2004) and
  Beers et al. (1999), respectively).  All comparison stars are hotter
  than Boo--1137, and would have slightly stronger Ca~II~K~3933\,{\AA}
  lines than seen here, if they were as cool as it.}

\end{figure}

\end{document}